\documentstyle[aaspp,12pt,flushrt]{article}

\begin{document}

\title{Discovery of Newly Formed Broad Absorption Lines in a Radio-Loud Quasar}

\author{Feng Ma}
\affil{Prc-Mrc 2nd Floor/R9950, University of Texas at Austin, 
Austin, TX 78712; feng@astro.as.utexas.edu}

\begin{abstract}

  We report a serendipitous discovery of broad absorption 
lines that were newly formed in the spectrum of the 
high-redshift, luminous radio-loud quasar TEX 1726$+$344, 
in a time interval of only 12 years. This is the first 
quasar showing a transition from narrow absorption lines 
to broad absorption lines. It also becomes one of the 
few radio-loud broad absorption line quasars. The gas 
cloud responsible for these broad absorption lines is 
derived to have parameters coinciding with those of 
the remnant of a tidally disrupted star. 

\end{abstract}


\section{Introduction}

 Approximately 10\% of all quasars are ``radio loud'', in the sense
that they are more luminous at radio
than at optical wavelengths. Also for 
some unknown reason,  about 10\% of radio-quiet quasars
show broad absorption lines (BAL) in their spectra. Only recently 
have a handful of radio-loud BAL
quasars been discovered (Becker et al. 1997; Brotherton et al. 1998), 
thus changing our 
view that BALs only appear in radio-quiet quasars. 
These quasars will serve as a link between radio-loud and 
radio-quiet quasars, and will aid in understanding the 
structure and kinematics
of powerful outflows in these quasars. This may eventually 
lead to a breakthrough
in understanding the origin of the radio loudness and 
the radio-loud/radio-quiet
dichotomy. 

\section{Observations}

In a spectroscopic survey of a sample of 62 radio-loud
quasars, we searched for emission line variations on a time
scale of $\sim$10 years (Ma \& Wills 1998; Ma 2000; Ma \& Wills 2001). 
A BAL system was serendipitously
discovered in the radio-loud quasar TEX 1726$+$344, 
which showed only narrow absorption lines just 12 years
ago. 
This quasar has 
a radio flux of 23.0 mJy or a 
luminosity of $10^{26.9}$ W Hz$^{-1}$ at 6 cm 
(Richards et al. 2001), a redshift of
2.43, and a V magnitude of 18.5 (Barthel, Tytler \& Thomson 1990).
The K-corrected ratio of radio-to-optical power $R^*$ (Sramek \& Weedman 1980; 
Stocke et al. 1992; Brotherton et al. 1998), 
commonly used as a measure of radio loudness, 
is derived to be $10^{2.36}$ and is well above the
criterion $R^*>10$ for radio-loud quasars. 

We observed  TEX 1726$+$344 on 2000 June 4, 
using the 2.7-m telescope at McDonald Observatory. 
A 2-arcsecond slit was used for a total exposure 
of 5000 seconds. The resolution is about 0.7 nm. 
The spectrum was corrected with 
a short exposure using an 8-arcsecond slit.
The corrected spectrum is denoted in the upper panel of 
Fig. 1 by a solid line. 
The historical spectrum taken by Barthel et al. (1990) 
 in 1988, after scaled to the continuum level of
the 2000 spectrum, is indicated by a dotted line. 
This historical spectrum was taken using the 
5-m Hale telescope with a 1-arcsecond slit in a 3600-second
exposure, and has a resolution of about 0.4 nm. 
The direct division spectrum and the fitting curve are shown in the lower panel. 

The continuum and the emission lines match well, as in most of 
the 62 quasars observed in a time interval of about 10 years. However, 
historical spectra of TEX 1726$+$344 in 1988 (Barthel et al. 1990) and 
also in 1990 (Lanzetta et al. 1991, not shown here)
show only narrow absorption lines. 
The BALs blueward the 
SiIV $\lambda$139.7 and CIV $\lambda$154.9 emission lines in the new
spectrum are dramatic, especially in the direct division spectrum. 

The SiIV 
and the CIV  absorption lines both have a 
redshift of 2.36, as compared to a
redshift of 2.43 in the emission lines. 
The difference in the peaks
between the absorption and the emission lines suggests that the absorbing
gas has an outflowing velocity of 6000 km/s.
Both absorption features are significant with restframe equivalent widths of 
0.3 nm and 0.2 nm, respectively. The full width at half maximum (FWHM)
of these BALs are approximately 1 nm 
in the restframe, corresponding to a
velocity dispersion over 2000 km/s. 

\section{Modeling}

The outflowing velocity and velocity dispersion coincide 
with those of a stellar remnant tidally disrupted by a
central massive black hole. It has been suggested that tidal disruptions 
may be responsible for quasar broad emission lines (Shields 1989; Roos 1992). 
A main sequence star, after being tidally disrupted by a $10^8M_{\odot}$ black 
hole in the center of a quasar, will be ejected at a high velocity. 
The head of the tidal debris will reach a distance of about 
$10^{17}$ cm in about 3 years, and the residual velocity 
will be 6000 km/s at this distance (Roos 1992). The thickness of the tidal 
debris is stretched at a velocity of  $\sim$1000 km/s, 
and the density drops to ${\sim}10^9$ cm$^{-3}$ at this point.  
The column density of the dense tidal stream is on the order 
of $10^{24}$ cm$^{-2}$ and is too dense to be responsible for 
the apparently un-saturated broad absorption lines in TEX 1726$+$344. 
However, as suggested by Ma (2000), the dense tidal stream is accompanied 
by faster-moving winds blown off of it due to radiation pressure. 
Both the density and the column density of the wind are two orders 
of magnitude smaller than those of the dense tidal stream. 

A model of this piece of gas cloud with a density of $10^7$ cm$^{-3}$ and  
a column density of $10^{22}$ cm$^{-2}$, 
at a distance of $10^{17}$ cm to a central continuum 
source with a luminosity of $10^{39}$ W, can thus be studied. 
We have run a photoionization calculation using {\small CLOUDY} (Ferland 1996), 
and found that the ratio of CIV to total carbon is $6.3{\times}10^{-5}$. 
Assuming a solar abundance with C:H$=3.55{\times}10^{-4}$ (Grevesse \& Anders 1989), 
we find that the column density of CIV of this piece of 
gas cloud is $2.2{\times}10^{14}$ cm$^{-2}$. 
This value coincides well with that derived from 
the equivalent width of the CIV broad absorption line.  

Column densities for the CIV and SiIV absorbing ions can be estimated by 
(Wang et al. 1999)
\begin{equation}
\displaystyle
N=\frac{W_\lambda}{\lambda}\frac{m_ec^2}{{\pi}e^2}\frac{1}{g{\lambda}f},
\end{equation}
where $W_\lambda$ and $\lambda$ are the equivalent width and the wavelength, 
and $g$ and $f$ are the effective statistical weight and oscillator strength
of the corresponding absorption lines, respectively. 
Following Wang et al. (1999) and taking the $g$ and $f$
values from Korista et al. (1991), we get the column density of CIV and SiIV to be
$2.4{\times}10^{14}$ and $7.3{\times}10^{13}$ cm$^{-2}$, respectively. 

\section{Discussions}

We note that 
absorption line variations with large magnitudes have been observed
in the low-redshift radio-quiet quasar PG 1126$-$041 (Wang et al. 1999).  
Numerous studies on the absorption line variability of 
Seyfert galaxies also exist in the literature 
(Malizia et al. 1997; Risaliti, Elvis \& Nicastro 2002; Akylas et al. 2002), 
and have been modeled with a single cloud moving through
the line of sight (Akylas et al. 2002). 
Hamann, Barlow \& Junkkarinen (1997) have identified 
variable intrinsic absorption lines (FWHM$\sim$400 km/s) 
in a radio-quiet quasar
Q2343$+$125. BAL variability at a moderate level ($<$40\%) has been
found in several radio-quiet quasars (Barlow 1993). 
In contrast to these previous studies, TEX 1726$+$344, 
a high luminosity {\it radio-loud} quasar
developing BALs in a time interval of 12 years, 
represents a new phenomenon. 

The popular disk-wind model (Murray et al. 1995) for 
both broad emission and absorption lines has been 
challenged by Ma \& Wills (2001) 
because some radio-loud quasars were discovered to have
dramatic CIV emission line variability, speculated to be due to 
the illumination of the jet pointing to the polar region of the quasar. 
The tidally disrupted stars model now seems to be the best candidate. 
The broad absorption lines in TEX 1726$+$344 will disappear in just 
a few years, if they were indeed caused by a tidal disruption event. 

\acknowledgements{ 
The author wishes to thank Bev Wills for advice;
Peter Barthel for making his data available in digital form;
 Gary Ferland for making his code {\small CLOUDY} available; and 
Angela Sinclair and Ariane Beck for help with the manuscript. 
This work makes uses of the NASA/IPAC Extragalactic Database (NED). 
}

\clearpage

\begin{figure}
\caption{Comparison of observed spectra at two 
epochs. The continua at different epochs have
been scaled to the same
level.   The solid line represents the new spectrum taken in 2000. The dotted
line is the  historical spectrum taken in 1988. The inset gives
expanded spectra in the CIV region. 
The subtraction spectrum is plotted in the same panel. In the lower
panel the direct division spectrum and the fitting curve are plotted together.}
\end{figure} 

\begin{figure}[hp]
\plotone{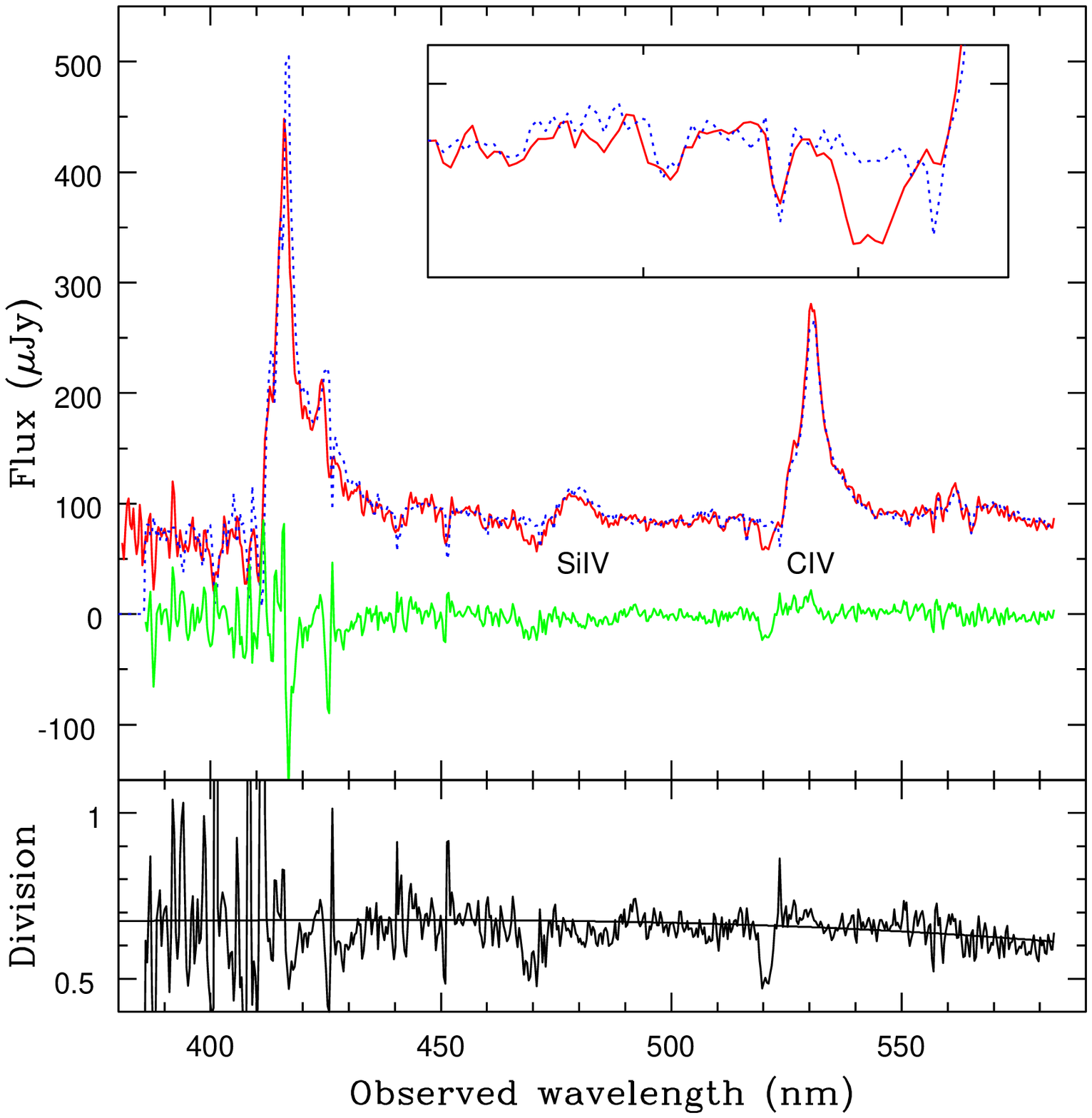}
\end{figure}


\begin{thebibliography}{}

\bibitem[]{} \reference Akylas A.,  Georgantopoulos I.,
 Griffiths R. G., Papadakis I. E.,
 Mastichiadis A., Warwick R. S.,
 Nandra K., Smith, D. A., 2002, 
MNRAS, 332, L23

\bibitem[]{} \reference Barlow T. A., 1993, PhD Thesis, 
Univ. California San Diego

\bibitem[]{} \reference Barthel P. D., Tytler D. R.,  Thomson B., 1990,
A\&AS, 82, 339

\bibitem[]{} \reference Becker R. H. et al., 1997,
ApJ, 479, L93

\bibitem[]{} \reference Brotherton M. S. et al., 1998, 
ApJ, 505, L7

\bibitem[]{} \reference Ferland G. J., 1996, 
Hazy, a Brief Introduction to {\small CLOUDY}, 
Univ. Kentucky Dept. Phys. Astron.
internal report 

\bibitem[]{} \reference Grevesse N., Anders  E., 1989, 
in Waddington C. J., ed, AIP Conf. Proc. 183, Cosmic Abundances 
of Matter, AIP, New York, p.1

\bibitem[]{} \reference Hamann F., Barlow T. A., Junkkarinen V.,
1997, ApJ, 478, 87

\bibitem[]{} \reference Korista K. T.,  et al., 1991, ApJ, 401, 529

\bibitem[]{} \reference Lanzetta K. M., et al., 1991, 
ApJS, 77, 1

\bibitem[]{} \reference Ma F., Wills B. J., 1998
ApJ, 504, L65

\bibitem[]{} \reference Ma F., 2000, 
PhD thesis, Univ. Texas Austin

\bibitem[]{} \reference Ma F., Wills B. J., 2001, Sci., 292, 2050

\bibitem[]{} \reference Malizia A., Bassani J., Malaguti G., 
Palumbo G. G. C., 1997, ApJS , 113, 311 

\bibitem[Murray et al. 1995]{murray95} 
\reference Murray N., 
Chiang J., Grossman S. A., Voit G. M., 1995, ApJ, 451, 498

\bibitem[]{} \reference Richards G. T., Laurent-Muehleisen S. A.,
Becker R. H., York D. G., 2001, ApJ, 547, 635

\bibitem[]{} \reference Risaliti G., Elvis M., Nicastro F., 
2002, ApJ, 571, 234

\bibitem[]{} \reference Roos N., 1992, ApJ, 385, 108

\bibitem[]{} \reference Shields G. A., 1989, in Osterbrock, D. E., Miller, 
J. S., eds, Proc. IAU Symp. 134, Active Galactic Nuclei, Kluwer Academic
                Publishers, Dordrecht, p. 577


\bibitem[]{} \reference Sramek R., Weedman D. 1980, ApJ, 238, 435

\bibitem[]{} \reference Stocke J. T., Morris S. L., Weymann R. J., 
Foltz C. B., 1992, ApJ, 396, 487

\bibitem[]{} \reference Wang T. G., Brinkmann W., Wamsteker W., Yuan W., 
Wang J. X., 1999, MNRAS, 307, 821


\end{thebibliography}
\end{document}